\documentclass[
reprint,
superscriptaddress,
amsmath,amssymb,
pre
]{revtex4-1}

\usepackage{color}
\usepackage{mathtools}
\usepackage{lipsum}
\usepackage{hyperref}
\usepackage{amsmath,amssymb}
\usepackage{graphicx}
\usepackage{dcolumn}
\usepackage{bm}
\usepackage{float}
\usepackage{epstopdf}
\usepackage{textcomp}
\usepackage[shortlabels]{enumitem}
\usepackage{verbatim}

\def\be{\begin{equation}}
\def\ee{\end{equation}}
\def\bea{\begin{eqnarray}}
\def\eea{\end{eqnarray}}
 
\def\nn{\nonumber}
\def\f{\frac}
\def\l{\left(}
\def\r{\right)}
\def\ls{\left[}
\def\rs{\right]}

\def\mr{\mathrm}
\def\refn{Eq.\,\ref}

\newcommand{\curie}{\affiliation{Laboratoire Physico Chimie Curie, Institut Curie, PSL Research University, CNRS UMR168, 75005 Paris, France.}}
\newcommand{\mpipks}{\affiliation{ Max Planck Institute for the Physics of Complex Systems, Dresden, 01187, Germany}}

\begin{document}
	
\title{Quantitative comparison of cell-cell detachment force in different experimental setups}

\author{Amit Singh Vishen}
\mpipks
\author{Jacques Prost}
\email{jacques.prost@curie.fr}
\curie
\author{Pierre Sens}
\email{pierre.sens@curie.fr}
\curie

\date{\today}
           
\begin{abstract}

We compare three different setups for measuring cell-cell adhesion. We show that the measured strength depends on the type of setup that is used.
For identical cells different assays measure different detachment forces. This can be understood from the fact that cell-cell detachment is  a global property of  the system. We also analyse the role of external force and line tension on contact angle and cell-cell detachment. Comparison with experiment suggests that the current data are essentially dynamical in nature.
We dedicate this article to Fyl Pincus who for many of us is an example to be followed not only for outstanding science but also for a marvelous human behavior.
\end{abstract}

\maketitle

\section{Introduction}

Integrity of a tissue is maintained by cell-cell adhesion
mediated by specific proteins \cite{Bell1984, Van2008}. The adhesion sites provide mechanical connectivity but also act as signaling hubs in the cell and are vital for tissue development and homeostasis. 
It has been shown that a differential expression of these adhesion proteins is essential for cell sorting \cite{Maitre2011,Maitre2012} and plays a central role in determining cell shape \cite{Hannezo2014}. 

Quantifying cell-cell adhesion is critical to understanding 
tissue mechanics. Significant experimental and theoretical progress have
been made in our understanding of the molecular processes underlying cell-cell adhesion. It has been long understood that adhesion mediated by specific linker proteins differs fundamentally from nonspecific adhesion
\cite{Bell1978, Bell1984, Bongrand1999}.
In the former case, the contact angle is set by the osmotic pressure difference of the mobile linkers between the adhesion zone and the plasma membrane \cite{Bell1984, Brochard2002}. Imaging of protein distribution on the surface has shed light on the spatio-temporal dynamics of the adhesion molecules that has further lead to a fruitful dialogue between theory and experiments on cell-cell adhesion.

A complementary approach to understanding adhesion has been to study cell-cell detachment. Different in-vitro assays have been designed to measure the response of a cell to shear force (due to fluid flow) or pulling (micropipette or a plate) with the aim to understand the parameters that determine the ability to cell to stay attached, either to another cell or to a substrate \cite{Kashef2015, Decave2002}.

The aim of this work is to synthesize these studies and highlight the differences in the measurement of detachment force for seemingly similar experimental setup and to compute the correlation between adhesion and detachment force. We compare three different setups used to measure the force required to detach two adherent cells: pipette-pipette \cite{Evans1985, Evans1985a, Evans1991, Chu2004, Simson1998}, plate-plate \cite{Puech2006, Desprat2005, Chaudhuri2009}, plate-pipette \cite{Prechtel2002} (Fig.~\ref{fig:schematic}). 

In this article we restrict our theoretical analysis to quasi-static conditions. This work follows closely Ref. \cite{Brochard-Wyart2003}. We analyze cell geometry, cell-cell detachment force, and force dependence of the contact angle. 
We provide analytical results for detachment force and cell shape in the pipette-pipette setup
and compare it with that for the plate-plate and plate-pipette setups. 
We show that the detachment force depends not only of the cell specific parameters like adhesion tension and cell surface tension but also strongly depends on the parameters of the experimental setup like pipette radius or cell-plate adhesion. For example, the detachment force for the same cells is always smaller when a pipette instead of a plate is used to hold the cells.

\section{Cell shape and  detachment force }

\begin{figure*}
	\centering
	\includegraphics[width= 0.7\linewidth]{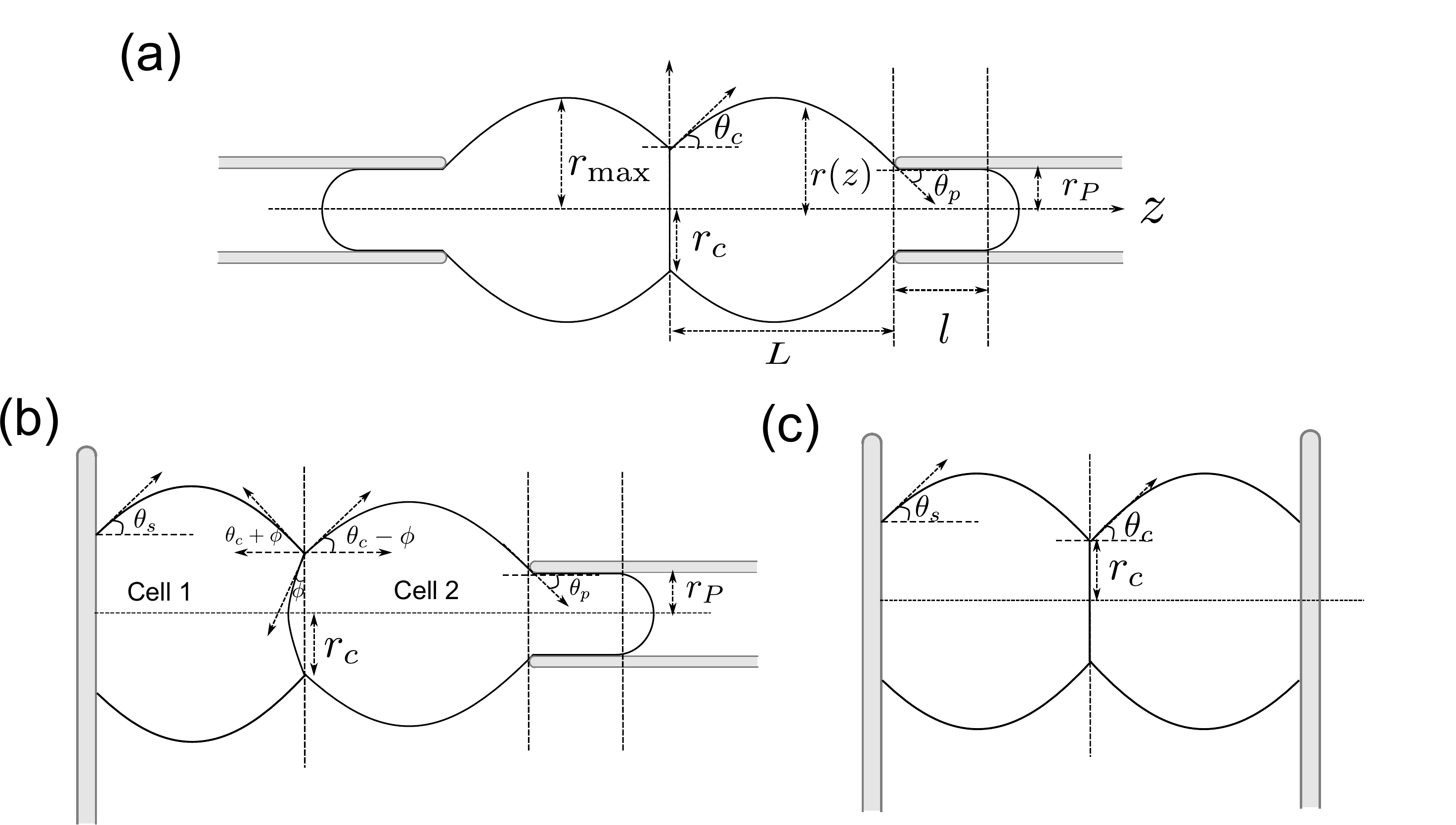}
	\caption{Schematic of the different experimental setups to measure cell-cell detachment: (a) cells pulled by pipettes (b) one cell on a substrate and the other pulled by a pipette (c) cells pulled by the substrates. 
	}
	\label{fig:schematic}
\end{figure*}

 We consider a doublet of identical cells adhering to each other. We take the cells to have a constant isotropic surface tension and a fixed volume.  The cell interior is treated as an incompressible fluid. The cells are also either sticking to a plate or sucked into a pipette, in such a way that the  three contact planes--cell-cell, cell-plate, or the cell-pipette--are parallel (with their normal along $z$-axis). Fig \ref{fig:schematic} shows the three setups that we analyze.

 At steady-state the stress in the cell interior, given by the hydrostatic pressure, and the cell tension can be taken to be 
 constant. As the problem is axis symmetric, the shape of the cell is completely specified by the function $r(z)$ (see Fig. \ref{fig:schematic}). For all $z$, the total force on the cross-section of the cell, perpendicular to $z$ axis should be equal to the external force, applied along $z$-axis, which reads \cite{Brochard-Wyart2003}: 
\be 
\label{eq:shape}
2 \pi r(z) \gamma \cos \theta(z) - \pi r^2(z) \Delta P_\mr{cell} = F,
\ee 
where  $r$ is the radius of the cell at $z$, $\gamma$ is the cell tension, $\Delta P_\mr{cell} = P_\mr{cell} - P_\mr{ext}$ is the pressure difference between the inside and outside of the cell,  $\theta$ is the angle between the tangent to the cell surface at $(r,z)$ and the $z-$axis, and $F$ is the external force applied on the cell. Positive value of $F$ implies pulling force on the cells. 
The force balance along the local normal of the cell is given by the Young-Laplace equation: $\Delta P = \gamma H$, where $H$ is the mean curvature, which is a constant.
Substituting this in \refn{eq:shape} we get:
\be 
\label{eq:shape1}
\,r^2 - \cos\theta\, r\,r_H + r_F\, r_H = 0,
\ee 
where we have defined $r_F = F/2\pi\gamma$ and $r_H = 2/H$. Note that both $r_F$ and $r_H$ can be negative. 
\refn{eq:shape1} gives us the full cell shape (see Appendix A). In Fig. \ref{fig:shapeforce} the shape obtained from \refn{eq:shape} is plotted for a given value of $r_F$ and $r_H$. 
For a given $\theta$, from \refn{eq:shape1} we get
\be   
\label{eq:radius}
r_\pm = \f{r_H\cos\theta  \pm \sqrt{r_H^2\cos^2\theta  - 4 r_F\, r_H}}{2}.
\ee   
 This solution is invariant under the change of sign of $\theta$.
 The roots are shown in Fig. \ref{fig:shapeforce} for a given value of $r_H$ and $r_F$. For a given radius there are two angles of equal magnitude and opposite sign, and for a given angle there are two radii.  The shape of the cell doublet is obtained by imposing the right boundary condition. For cell-cell contact this is determined by the force balance in the contact plane. For the simple case of a constant adhesion energy this imposes a constant contact angle given by:
 \be 
 \label{eq:contact_angle}
 \gamma \sin \theta_c = \gamma - w,
 \ee 
where $w$ is the adhesion tension and $\theta_c$ is the contact angle. The effect of external force on contact angle is discussed in more details below. The contact radius is given by \refn{eq:radius} with $\theta = \theta_c$, this implies we need to further specify which of the two solutions is the contact radius at the boundary. On the other side the boundary condition is given by the size of the pipette, or by the adhesion strength of the cell-plate contact.   

The roots in \refn{eq:shape1} are real when $r_H\cos^2\theta > 4\,r_F$. For $r_F > r_H/2$, there is no solutions to the shape equation. When there is an angle imposed at the boundary, like contact angle due to cell-cell adhesion, then the value of maximum force is $r_F = r_H\cos^2\theta_c/4$ \cite{Brochard-Wyart2003}. Above this value of force, one of the assumptions of the model breaks down. 
If the tension remains constant then the cell needs to detach. 
This relation was first analyzed in Ref.~\cite{Brochard-Wyart2003}, where the detachment force is characterized in terms of the contact radius and the maximum radius. 

The contact radius at detachment is $r_H\cos\theta_c/2$. The force needed to detach the cell-cell contact reads:
\be  
\label{eq:FDC}
F_\mr{dc} = \f{1}{2}\pi \gamma\,r_H \cos^2\theta_c. 
\ee  
Note that, depending on the setup, both $r_H$ and $\theta_c$ can be a function of $r_F$. 

\begin{figure}
	\centering
	\includegraphics[width= 0.9\linewidth]{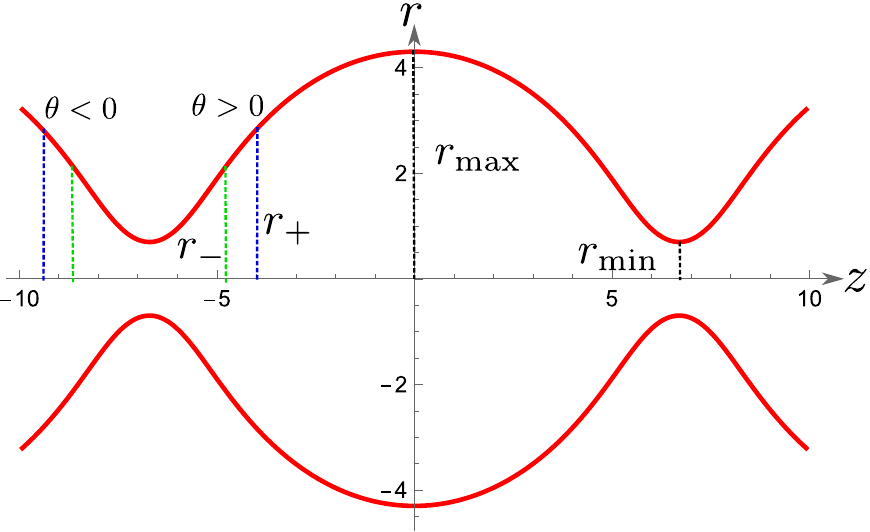}
	\caption{Shape of constant curvature obtained by solving \refn{eq:shape} for $r_H = 5~\mu m$ and $r_F = 0.6~\mu m$. The maximum and minimum radius are given by \refn{eq:radius} for $\theta = 0$. The plot shows the two radii $r_\pm$ corresponding to $\theta = \pi/6$. 
	}
	\label{fig:shapeforce}
\end{figure}

\section{Detachment force for micropipette setup}

\begin{figure}
	\centering
	\includegraphics[width= \linewidth]{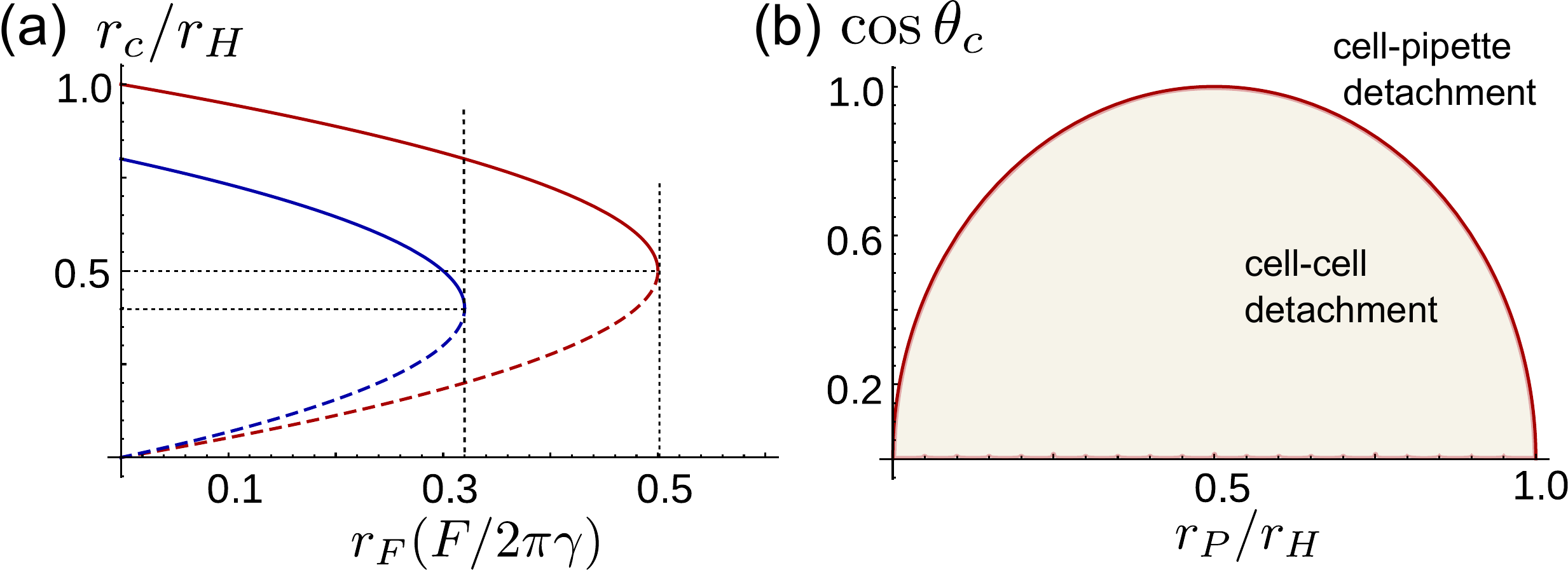}
	\caption{(a) The plot shows the contact radii  as a function of external force obtained from \refn{eq:radius} for $\theta = \theta_c$, where the two solutions are show by the dashed and dotted part of the curve. The red plot is for $\theta_c = 0$ and blue is for $\cos\theta_c = 0.8$, for both plots $r_H = 2$. (b) The plot shows the region for which the plot showing the region for cell-cell and cell pipette detachment as a function of contact angle and pipette radius, for $r_H = 2$.}
	\label{fig:force_radius}
\end{figure}

We now analyze, in detail, the case of two adhered cells pulled apart by pipettes Fig. \ref{fig:schematic}(a)). For a cell to be held in a pipette we need $\Delta P_\mr{tube}<0$ where $\Delta P_\mr{tube}$ is the pressure difference between the inside and the outside of the pipette. 
To know the pressure inside the cell we need to know the shape of the cell  in the pipette, the cell tension in the pipette, and boundary conditions between the cell and the pipette. For simplicity and since in experiments careful coating ensures the absence of adhesion, we ignore any adhesion between the cell and the pipette. As a result, the extremity of the tongue inside the pipette is a hemisphere, which corresponds to zero contact angle.
Furthermore, we consider quasi-static situations and we take the cell tension to be uniform inside and outside the pipette. 

The external force applied along the z axis on a plane perpendicular to the pipette is 
\be 
\label{eq:force-ext}
F = F_p - \pi r_P^2\,\Delta P_\mr{tube},
\ee 
where $F_p$ is the external force applied on the pipette, and $r_P$ is the radius of the pipette. 
Using the Young-Laplace equation inside the pipette we get $\Delta P_\mr{cell} - \Delta P_\mr{tube} = 2\gamma/r_P$.  There is an important observation to be made here, for a cell of a given tension, the pressure inside the cell is fixed by the pressure and the radius of the pipette. 
\be  
\label{eq:laplace_pipette}
\f{2\gamma}{r_H} = \Delta P_\mr{tube} + \f{2\gamma}{r_P}.
\ee   
We see that $r_H$ and $r_P$ are not independent, the mean curvature of the cell can be varied by changing the pipette radius or the pressure in the pipette.  From \refn{eq:force-ext} and \refn{eq:laplace_pipette} we see that changing $F_p$ changes $r_F$ but it does not lead to a change in $r_H$. 
However, changing $\Delta P_\mr{tube}$ leads to a change in both $r_F$ and $r_H$.

Since the radius of the pipette is fixed, the angle between the pipette and the cell at cell-pipette contact changes in response to the applied force.  \refn{eq:shape1} evaluated at $r=r_P$ gives the angle at the pipette, which reads
\be 
\label{eq:angle_pipette}
\cos\theta_p = \f{(r_P^2 + r_Hr_F)}{r_P r_H}.
\ee 
We define the force for which $\theta_p = 0$ as the cell-pipette detachment force. 
Substituting \refn{eq:laplace_pipette} in \refn{eq:angle_pipette}, the detachment force for cell-pipette contact reads:
\be
F_\mr{dp} = -\pi r_P^2 \Delta P_\mr{tube}.
\ee 
Thus we see that the cell-pipette detachment force depends on the pipette pressure and size. 
Substituting the value of $r_H$ from \refn{eq:laplace_pipette} in \refn{eq:FDC} the cell-cell detachment force reads:
\be 
\label{eq:force_detach_1}
F_\mr{dc} = \f{\pi r_P \gamma^2 \cos^2\theta_c}{2\gamma + r_P \Delta P_\mr{tube}}.
\ee  
The cell-cell detachment depends not only on the adhesion strength and cell tension (that defines $\theta_c$) but also on the pipette size and the pipette pressure. 

Whether it is  the cell-cell contact or the cell-pipette contact that detaches first depends on the size of the pipette and the 	contact angle. If $F_\mr{dc} < F_\mr{dp}$ then the cell-cell contact detaches before the cell-pipette contact. 
Fig \ref{fig:force_radius} (b) shows the region in which cell-cell detachment happens before cell-pipette detachment for different contact angles and pipette radius. The ratio of the cell-cell contact size at detachment and for zero force,  as obtained from \refn{eq:shape1} is exactly half. 

\section{Comparison between micropipette and microplate setups}

In the microplate setups the cells are adhered to a plate \cite{Yoneda1964, Yoneda1986, Yang2017, Yang2018} (Fig. \ref{fig:schematic}(b)).
Superficially the microplate and the micropipette setups look similar, in both cases the cell shape is given by \refn{eq:shape1} and the detachment criterion is given by \refn{eq:FDC}. However, there is one fundamental  difference, the value of $r_H$ in the cell-pipette setup is set by the pipette radius and pipette pressure (\refn{eq:laplace_pipette}), whereas $r_H$ in the cell-plate setup needs to be determined self-consistently from the constraint on the cell volume. 
One can even think of the micropipette setup as a constant pressure ensemble and the microplate setup as a constant volume ensemble. 

\begin{figure}
	\centering
	\includegraphics[width= \linewidth]{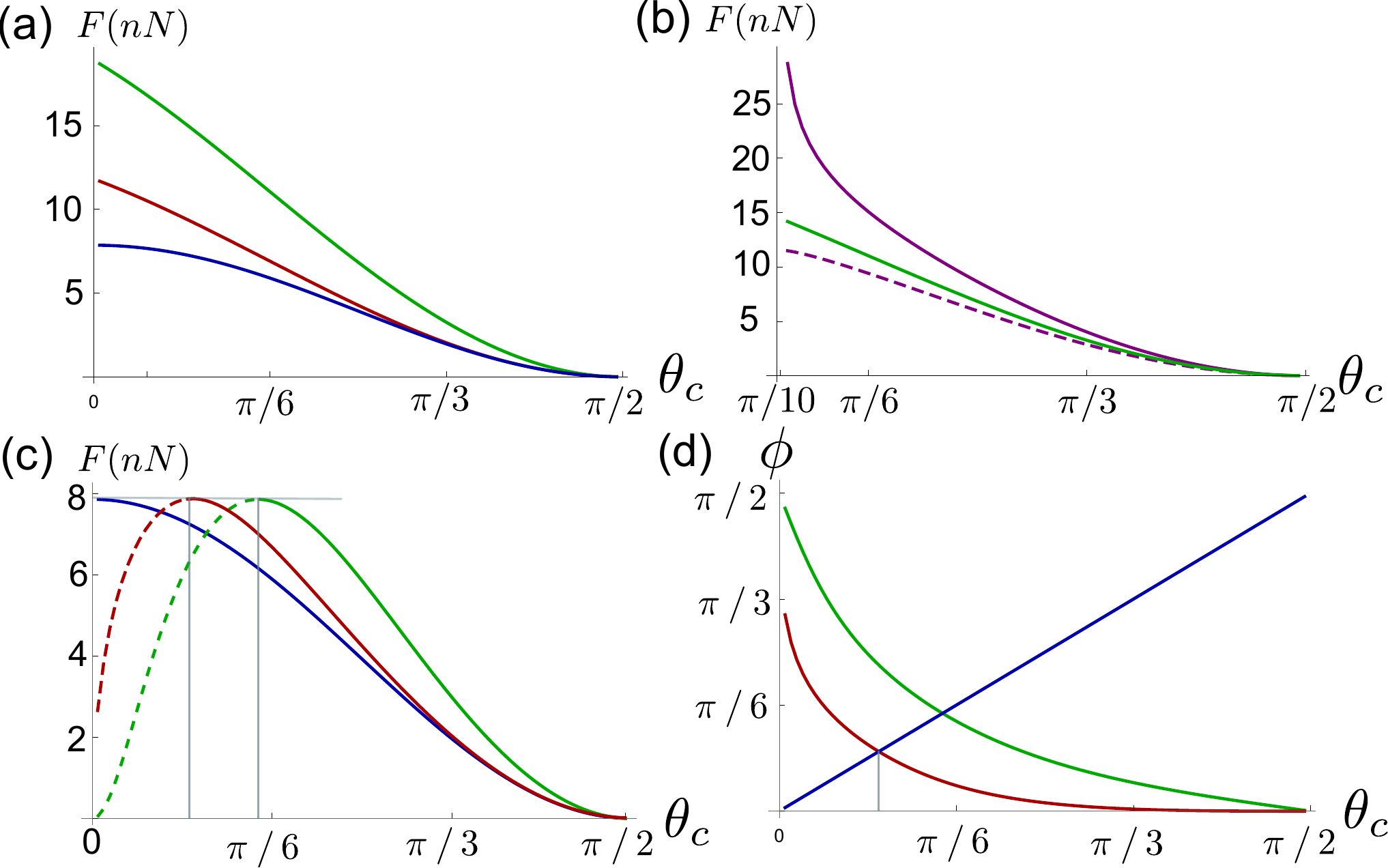}
	\caption{(a) Detachment force as a function of contact angle. The blue curve is for micropipette setup for $r_H = 5~\mu m $, and the red and green curves are for the microplate setup with zero cell-plate contact angle, for initial cell radius of $5~\mu m$ (red) and $8~\mu m$ (green).
 (b) Plot showing a comparison between detachment forces for cell-plate contact angle of $-\pi/10$ (purple solid line),$\pi/10$ (purple dashed line), and zero (green curve). The cell size cell is  $8~\mu m$. (c) Detachment force in plate-pipette setup. As in the previous plot the blue curve corresponds to the micropipette setup, the green curve is for the cell size  $8~\mu m$ and the red curve is for $5~\mu m$. (d) The angle $\phi$ at detachment as a function of contact angle. For force calculation we have taken $\gamma = 1~mN/m $.
 }
	\label{fig:detachmentforce}
\end{figure}

Fig. \ref{fig:detachmentforce} (a) shows the comparison between the detachment force for the same cells in pipette-pipette and plate-plate setup. For a given contact angle, the detachment force measured by the latter is consistently higher than the detachment force measured by the former setup. This can be understood by noting that the value of $r_H$ is larger in the case of plate-plate setup than that in the case of pipette-pipette setup, and from \refn{eq:FDC} we see that the detachment force is proportional to $r_H$.

In the cell-plate setup the detachment force depends on the contact angle, cell tension, cell volume, and cell-plate adhesion, whereas in the cell-pipette setup the detachment force depends on the contact angle, cell tension, pipette radius, and pipette pressure. Fig. \ref{fig:detachmentforce} (a) shows the effect of cell volume on the detachment force. The bigger the cell, the smaller is the hydrostatic pressure difference(larger $r_H$); hence, bigger is the detachment force.  \ref{fig:detachmentforce} (b) shows the effect of the cell-plate contact angle on the cell-cell detachment force. The detachment force is larger for smaller contact angle. This dependence of detachment force on the cell-plate contact angle can be understood from the inverse relation between the cell-plate contact angle and $r_H$ for a given cell-cell contact angle. We also note that for the micropipette setup the detachment force is the same for positive and negative $\theta_c$. However, for the microplate setup it is not the case. This is because the pressure in the cell is not the same for the positive and negative value of $\theta_c$; the pressure in the cells with negative $\theta_c$ is smaller than that of the cells with positive $\theta_c$.

We now discuss the case when one cell is pulled by a pipette while the other 
is attached to a substrate (Fig. \ref{fig:schematic}(c)). When the cell plate contact angle is zero, the plate-pipette setup also corresponds to the cell detachment experiments on cell-triplets \cite{Maitre2012}. This system is asymmetric even if the two cells are identical.  For identical volume and tension the two cells have different hydrostatic pressures and as a result 
the cell-cell interface is curved. The radius of curvature of the interface is given by $r_I = 2\gamma_c/(P_2 - P_1)$,
where $\gamma_c=\gamma-w$ is the tension in the adhesion plane, $P_1$ and $P_2$ are the hydrostatic pressures in the cells attached to plate and pipette, respectively. 
From the geometry we get $r_c/r_I = \sin \phi$ (see Fig.~ \ref{fig:schematic}(b)). 
For simplicity we take the surface tension of the two cells to be the same. Substituting $\Delta P_2 = 2\gamma/r_{H_2}$, $ \Delta P_1 = 2\gamma/r_{H_1}$, and $\gamma_c = \gamma\sin\theta_c$ we get: 
$r_I  = r_c/\sin\phi$.
We want to compute the angle $\phi$ at the point of cell-cell detachment. The effective contact angle for the cell attached to the pipette is $\theta_c - \phi$. The detachment condition gives $r_c = r_{H_2}\cos\l \theta_c - \phi\r/2 $, and the detachment force is $r_F = r_{H_2} \cos^2\l \theta_c - \phi\r/4$. Substituting $r_c$ we get
after few steps of algebra 
$r_{H_1}/r_{H_2} = \cos\l \theta_c - \phi\r/\cos\l \theta_c + \phi\r $.
Substituting $r_F$ and $r_{H_1}$ in term of $\phi$ in the volume conservation equation of  the cell attached to the microplate we obtain $\phi$. 

Fig. \ref{fig:detachmentforce} (c) shows the comparison between the detachment force for the same cells in pipette-pipette and plate-pipette setups for two different cell sizes. We  see that the detachment force is consistently larger for the plate-pipette setup when $\theta_c >\phi $ (bold part of the curve) and just like the microplate setup the lager cell has larger detachment force. 
Fig. \ref{fig:detachmentforce} (d) shows the $\phi$ as a function of contact angle for two different cell sizes. The dotted part of the plot in Fig. \ref{fig:detachmentforce} (c) corresponds to the value of $\theta_c$ for which $\theta_c < \phi$ at detachment, i.e, the effective contact angle is less than zero. For these values of contact angle the detachment force for the plate-pipette can be smaller than that of the pipette-pipette.

To summarise, the detachment force measured by the pipette-pipette setup is smaller than that in the plate-plate setups. The difference decreases with increasing contact angle. The detachment force for the plate-pipette setup as a function of contact angle is non-monotonic, it can be smaller than the pipette-pipette detachment force for contact angles smaller than a threshold.  This result shows the subtlety involved in interpreting the detachment force as a proxy of adhesion strength.

\section{Force dependence of contact angle}

The above analysis shows that for the same contact angle different experimental setups measure different detachment force.
We now take a closer look at what sets the contact angle and whether taking the contact angle to be constant under changing external force is justified. 

A prevailing idea is that when the cells are pulled apart the adhesion molecules are stretched and are more likely to detach than they are in the absence of force. Since the adhesion strength is determined by the number of adhered molecules the adhesion tension then decreases with increase in force \cite{Bell1978, Yang2017}. Although this sounds reasonable, one flaw in this argument is that the applied force is not directly felt by the adhesion molecules. 

Here we present a simple model of adhesion which includes linkers that can bind and unbind. When the linkers can freely diffuse on the cell surface, the adhesion leads to a decrease in cell tension given by the two dimensional osmotic pressure difference between the contact plane and the free cell surface \cite{Bell1984, Brochard2002}. In the simple case of dilute bound linkers the osmotic pressure is $k_B T n_b$, where $n_b$ is the density of bound linkers \cite{Bell1984, Brochard2002}. Substituting $w = k_B T n_b$ in \refn{eq:contact_angle} gives us the contact angle. 

For quasistatic pulling rates, the density of bound linkers is given by a Boltzmann distribution
$n_b \propto n_{b0} e^{-\beta k (l-l_0)^2/2}$,
where $\beta =1/(k_BT)$,$k_B$  is the Bolztmann constant, $l$ the length of the linkers, $l_0$ the length at zero force, and $k$ the stiffness. In practice the quasistatic pulling rate is obtained when it is much slower than the time it takes for the linker density to reach equilibrium.
Away from the periphery of the adhesion zone the membrane is flat, the only force on the linkers is the hydrostatic pressure difference between the cell and the adhesion zone. 
The force balance reads $-k n_{b0} (l - l_0) = \Delta P$ which gives
\be  
\label{eq:bound-density}
n_b = n_{b0} e^{-\beta \Delta P^2/2 k n_{b0}^2}.
\ee 
Substituting in \refn{eq:contact_angle} $w = k_B T n_b$ 
and using \refn{eq:bound-density} the contact angle reads
\be 
\sin\theta_c = 1 - \f{k_B T n_{b0}}{\gamma}e^{-\beta \Delta P^2/2 k n_{b0}^2}.
\ee 
For the micropipette experiment the value of $\Delta P$ can be maintained constant while changing the applied force. In this case the contact angle is constant but it depends on the experimental condition.  For the microplate experiment the value of the contact angle indeed depends on the applied force as the cell pressure depends on it. The limit $\Delta P \to 0$ gives us the upper limit of the detachment force and $\Delta P = 2\gamma/r_0$, where $r_0$ is the radius of the free cell, gives us the lower limit on the detachment force. 

For $k \approx 1\,pN/\mu m = 10^{-5} N/m$, $\Delta P \approx 100\,Pa$, $n_{b0} \approx 1000/\mu m^2 = 10^{15}/m^2$, $\beta \approx 2\times10^{20} J^{-1}$ we get $\beta\Delta P^2/2k n_{b0}^2 \approx 1$. 
This shows that the effect of hydrostatic pressure difference on adhesion strength can be appreciable for physiologically relevant parameter values.

\section{Detachment force with line tension}

A clever experiment allowing to visualize the distribution of E-cadherin, actin and myosin at the cell-cell interface, shows that under quasistatic conditions, these three molecules are essentially expelled from the bulk of the contact region and concentrated in a ring with an actin-myosin distribution reminiscent of stress fibers \cite{Engl2014}. This structure is very likely to be contractile. To include its effect on the adhesion mechanics we add a line tension term in the \refn{eq:contact_angle}. The force balance in the direction tangent to the contact place now reads
\be
\label{eq:contactangle_withline-tension}
\gamma \sin\theta_c = \gamma - w + \f{\gamma_l}{r_c}, 
\ee 
where $\gamma_l $ is the line tension, which can be positive or negative. Positive line tension leads to a weaker adhesion compare to that without it.
Substituting the contact angle obtained from \refn{eq:contactangle_withline-tension} in \refn{eq:shape1} we get
\be 
x^4 + \l \f{2 r_F}{r_H}  - \cos^2\theta_{0} \r x^2 + \f{2 \sin\theta_0 r_l}{r_H} x +  \f{r_F^2}{r_H^2} + \f{r_l^2}{r_H^2}  =0, 
\ee 
where we have defined $x = r_c/r_H$,$\bar \gamma = 1- w/\gamma$, and $r_l = \gamma_l/\gamma$.

\begin{figure}
	\centering
	\includegraphics[width= 0.9\linewidth]{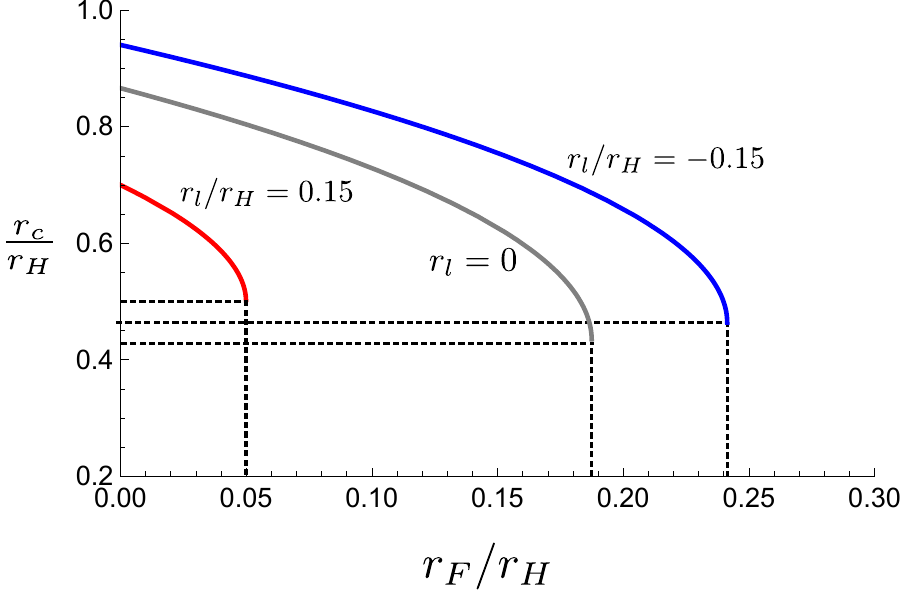}
	\caption{Plot showing the contact radius as a function of force for different values of line tnesion: $r_l=0$ (gray), $r_l/r_H = 0.15$ (red), and $r_l/r_H = -0.15$ (blue). For all the plots $\theta_0 = \pi/6$. The dotted lines mark the contact radius and force at detachment. 
	}
	\label{fig:line_tension}
\end{figure}

From Fig. \ref{fig:line_tension} we see that for positive line tension the cells detach at a smaller force and smaller contact radius in comparison to the cell without line tension. For the particular choice of $\theta_0$ of Fig. \ref{fig:line_tension} the relative reduction in contact radius is only of a few percent whereas the detachment force has changed by a large factor. 
$r_l/r_H = \gamma_c d/\gamma r_H$, where $\gamma_c$ is tension in the contractile ring that is effectively treated as line tension and $d$ is the width of the ring. For $\gamma_c \approx 1.5 \gamma$, $r_H = 5\,\mu m $ and $d = 500\,\mr{nm}$ we get $r_l/r_H \approx 0.15$ that is used to plot Fig. \ref{fig:line_tension}.

\section{Discussion}

The main conclusion of this work is that the different experimental setups discussed in this article measure different detachment forces for the same doublet of cells, and in no case measure an adhesion energy. The detachment force measured in the microplate setup is always larger than that measured in the micropipette one, whereas in the mixed case the detachment force has an intermediate value. We can infer the adhesion tension ($w$ in \refn{eq:contact_angle}) in all cases if we know some of the geometric and mechanical parameters related to the cell and the setup. To compute the adhesion tension  in the pipette-pipette and plate-plate cases the knowledge of different combination of parameters is required. The detachment force in the former case, depends on adhesion tension, surface tension, pipette radius, and hydrostatic pressure in the tube. Thus to infer adhesion tension the only cell parameter needed is cell tension. For the plate-plate setup we need to know the cell volume as well as the cell tension to infer the adhesion tension.
One must keep in mind though, that $w$ is not the adhesion energy but rather the 2D osmotic pressure of adhesion molecules.

Although generally the focus of the experiments is on adhesion tension, the detachment force is an important physiological parameter in its own right. These two quantities are related but can be independently modulated. 

Our analysis is valid for circularly symmetric geometries and includes cases where the cortex in the contact zone is strongly depleted, except in a narrow zone close to the edge of the cell-cell contact. In this case the line tension introduced in \refn{eq:contactangle_withline-tension} plays an important role. Cases, with phase separated domains would require a different analysis \cite{Bruinsma2001}.

In this work we consider pulling rates slow enough 
that the cell tension can be taken to be  constant. For a finite pulling rate the cortical tension can be approximated as the sum of a static part given essentially by the cell cortex contractility $\gamma_\mr{ss}$ and a dynamical term due to cortex viscosity $\gamma_\mr{dynamic}$, that is $\gamma(t) = \gamma_\mr{ss} + \gamma_\mr{dynamic}$, where 
$\gamma_\mr{dynamic} \approx \eta h\, dA/A dt$, where $\eta$ is 3d viscosity of the cortex, $h$ is cortical thickness, and $A$ is the cell surface area. We can approximate $ dA/Adt \approx 1/\tau$, where $\tau$ is the timescale of pulling. Taking $\gamma_\mr{ss} \approx \zeta h$ with contractility of the cortex $\zeta \approx 5~kPa$, $h \approx 200 ~nm$ we get $\gamma_{ss} \approx 1~mN/m$. For $\eta \approx 10^5~Pa.s$ we get $\gamma_\mr{dynamic} \approx 20/\tau~mN/m$. For strong adhesion the detachment force is  $ \pi\gamma r_H/2$ 
which leads for slow pulling rates
detachment forces on the order of $10~nN$ for $r_H$ of few micrometers. Such detachment forces are consistent with some
measurements
\cite{Maitre2012}. 
Note that the dynamics of the linkers of the cortex in the contact place can only lead to a lower detachment force not higher than obtained in the limit $\theta_c = 0$. A detachment force of the order of $100~nN$, as noted in other
experiments \cite{Chu2004},  
implies that there is a significant viscous contribution to the stress. 
A complete dynamic model is outside the scope of this work but could be studied using the framework of active surfaces \cite{Berthoumieux2014, Mietke2019}.

\section{Acknowledgement}

Saint Pierre and le Roi des cons thank Fyl Pincus for illuminating and friendly discussions which inspired them more than once.
ASV received support from the grants ANR-11-LABX-0038, ANR-10-IDEX-0001-02. 
The authors thank Virgile Viasnoff for useful discussions.
\\

\noindent{\bf Author contribution statement:}
ASV, JP, and PS designed the study. ASV, under the supervision of JP and PS, performed the theoretical and numerical analysis. ASV, JP and PS wrote the manuscript.\\

\noindent{\bf Data availability statement:} we do not analyse or generate any datasets, because our work proceeds within a theoretical and mathematical approach.

\appendix

\section{Shape for constant curvature}{\label{appendix_shape}}

We can obtain the shape of the cell form \refn{eq:shape} \cite{Yoneda1964, Fischer2014}.
The shape is of constant curvature surface shown in Fig. \ref{fig:shapeforce}. The shape is axissymmetric, we parametrize it by arc length $s$ starting from $s_c$ at the cell-cell contact to $s_p$ at the cell-pipette contact. Using \refn{eq:shape} and $d r/d s = \sin\theta$ we get
\be 
\f{d r}{d s} = \pm \sqrt{1 - \l \f{r}{r_H} + \f{r_F}{r}\r^2}.
\ee 
The positive and negative sign corresponds to positive and negative $\theta$, respectively. After integration this gives
\be 
\label{eq:arclength}
\pm \f{2 s}{r_H} = -\arccos \ls \f{ (r^2 - r_\mr{max}^2) + (r^2 - r_\mr{min}^2)}{(r^2_\mr{max} - r^2_\mr{min})}\rs,
\ee
where $r_\mr{max} = (r_H +\sqrt{r_H^2 - 4 r_H r_F})/2$ and $r_\mr{min} = (r_H - \sqrt{r_H^2 - 4 r_H r_F})/2$. Note that $r_H = r_\mr{max} + r_\mr{min} $. The constant of integration is chosen such that $s=0$ corresponds to $r=r_\mr{max}$.
\be 
\label{eq:radius1}
 r(s) =r_\mr{max} \sqrt{1 - \alpha \sin^2\l \f{ s}{r_H}\r}, 
\ee 
where $\alpha = 1 - r^2_\mr{min}/r^2_\mr{max}$. Using \refn{eq:shape} and $d z/d s = \cos\theta$, we get 
\be 
\f{dz}{ds} =  \l \f{r}{r_H} + \f{r_F}{r}\r,
\ee 
Upon integration this gives
\be
\label{eq:position}
 z(s) = r_H \ls \f{r_\mr{max}}{r_H} E_2 \l \mp\f{s}{r_H}, \alpha \r + \f{r_F}{r_\mr{max}} E_1 \l \mp\f{s}{r_H}, \alpha\r\rs,
\ee 
where $E_1$ and $E_2$ are incomplete elliptic integrals of the first and second kind, respectively. We have set the origin of $z$-axis at $s = 0$.
Assuming the pipette and the cell-cell contact to line on the opposite side of the maximum and within the two minimum, the distance between the cell-cell contact and the pipette is  $L = z(-s_c) + z(-s_p)$, where $s_c$ and $s_p$ are obtained from \ref{eq:arclength} for $r = r_c$ and $r= r_P$, respectively. The separation between the pipettes is given by
\bea
\label{eq:length}
\nn 2L &=& -2 r_H \l \f{r_\mr{max}}{r_H} \ls E_2 \l \f{s_p}{r_H} , \alpha\r + E_2 \l \f{s_c}{r_H} , \alpha\r\rs \right.\\
&+&  \left. \f{r_F}{r_\mr{max}} \ls E_1 \l \f{s_p}{r_H}, \alpha\r + E_1 \l \f{s_c}{r_H}, \alpha\r\rs \r .
\eea 
The negative sign is obtained using the relation $E_{1,2}(s,\alpha) = - E_{1,2}(-s,\alpha)$.

The surface area and the volume of the cell can also be obtained in term of elliptic integrals \cite{Yang2017}. Note \refn{eq:length} is the expression of length for both plate and the pipette setup. Since $r_H$ is know for the pipette setup this equation gives us the separation for a given force. However, for the plate setup, we need an additional equation to compute $r_H$. A constraint of volume provides this additional constraint that can be used to compute $r_H$. 

The surface area of the cell outside the pipette is given by 
\be
A = \int_0^{s_p}2\pi r(s) ds + \int_0^{s_c} 2\pi r(s) ds.
\ee
Substituting $r$ and integrating we get:
\be
A =  -2\pi r_H r_\mr{max} \ls E_2 \l \f{s_p}{r_H} , \alpha\r + E_2 \l \f{s_c}{r_H}, \alpha\r\rs.
\ee
The volume of the cell outside the pipette is given by
\be
V = \int_0^{z_p}\pi r^2 dz + \int_0^{z_c} \pi r^2 dz.
\ee

\bibliographystyle{apsrev4-1}
\bibliography{CellCellDetachment}{}

\end{document}